\documentclass[prb,aps,preprint]{revtex4}
\usepackage{graphicx}
\usepackage{amsfonts}
\usepackage{amssymb}
\usepackage{amsmath}
\usepackage{textcomp}

\usepackage{color}
\usepackage{psfrag}

\usepackage{subfigure}

\usepackage{epsfig}

\begin{document}

\title{ Designed Diamond Ground State via Optimized Isotropic Monotonic Pair Potentials }


\author{{\' E}. Marcotte,$^{1}$  F. H. Stillinger, $^{2}$ and Salvatore Torquato$^{1,2,3,4}$}
\email{torquato@princeton.edu}
\affiliation{$^{1}$Department of Physics,
$^{2}$Department of Chemistry, $^{3}$ Program in Applied and Computational Mathematics, $^{4}$ Princeton Institute of the Science and Technology of Materials, 
Princeton University, Princeton, New Jersey 08544, USA}

\begin{abstract}

We apply inverse statistical-mechanical methods to find a simple family of optimized isotropic,
monotonic pair potentials, under certain constraints,
whose ground states for a wide range of pressures is the diamond crystal.
These constraints include desirable phonon spectra and the widest possible
pressure range for stability.
We also ascertain the ground-state phase diagram
for a specific optimized potential to show that
other crystal structures arise for other pressures. Cooling disordered
configurations interacting with our optimized potential to absolute zero 
frequently leads to the desired diamond crystal ground
state, revealing that the capture basin for the global energy minimum
is large and broad relative to the local energy minima basins.
\end{abstract}

\maketitle


Advances in the field of self-assembly, devising building blocks
(\emph{e.g.},  nanoparticles and polymer chains)
with specific interactions to form larger functioning materials, are proceeding rapidly and hold
great promise to produce unique colloidal and polymer systems. \cite{Gr10, Kr11,Fr11,Sh11}
In the past several years, inverse statistical-mechanical
methods have been formulated that yield optimized interactions that
robustly and spontaneously lead to a {\it targeted} many-particle
configuration with desirable or novel bulk properties. \cite{To09}
This inverse approach provides a powerful and systematic means of directing
self assembly with exquisite control. Recent studies
have used inverse methods to find optimized isotropic
(non-directional) interactions, subject to certain constraints, that yield novel targeted 
ground states, such as low-coordinated crystal structures. \cite{To09}
This includes the three-fold coordinated honeycomb (or graphene) structure in two dimensions 
\cite{Re05} 
and the tetrahedrally-coordinated diamond crystal in three dimensions,\cite{Re07} initial
studies  of which involved isotropic pair potentials with multiple wells.

Are multiple wells required to achieve low-coordinated crystal ground
states with isotropic pair interactions? We have recently shown that inverse 
statistical-mechanical techniques allow one to produce robustly both the square
lattice and honeycomb crystal in two dimensions via monotonic convex pair potentials.\cite{Ma11a,Ma11b} 
In the present work, we employ inverse techniques 
to obtain  a simple family of optimized isotropic, monotonic pair potentials whose ground states for a 
wide range of pressures is the diamond crystal.

Using the {\it forward} approach,\cite{forward} it was established over a decade ago that
the diamond crystal can be stabilized for a range
of densities by an isotropic, monotonic pair potential devised to model star polymers. \cite{Wa99}
These authors used free-energy calculations to find the phase
diagram and validate their conclusions. Moreover, the potential possessed stable phonon spectra over the predicted
ground-state parameter regime.\cite{private}
A forward approach was used in another study \cite{Pr09}
to examine only lattice energy sums at zero temperature for a relatively
small set of Bravais and non-Bravais lattices for an isotropic, monotonic pair potential. It
was found that the diamond crystal was stable for a certain pressure range.
These authors recognized the limitations of this restricted
investigation, which excluded both phonon spectra calculations
and annealings to zero temperature from liquid-like
initial conditions in order to validate that
the diamond was indeed the ground state.

Here we use a simpler functional form for a monotonic radial (isotropic)
pair potential function $v(r)$ that obeys certain important conditions on the
second derivative with respect to the radial interparticle distance $r$
established in Refs. \onlinecite{Ma11a} and \onlinecite{Ma11b}.
Specifically, we propose a potential function of the form
\begin{equation}
v(r) = \varepsilon \left(1 + a_1 \frac{r}{\sigma} + a_2 \left(\frac{r}{\sigma}\right)^2 \right) e^{-\left(r / \sigma\right)^2}, \label{eqn:potential}
\end{equation}
where $\varepsilon$ and $\sigma$, respectively, define the energy and length
units, and $a_1$ and $a_2$ are dimensionless parameters.
Equation~(\ref{eqn:potential}) is chosen for its simplicity
and because it allows for the desirable features of the second derivative
described below. 
The potential function (\ref{eqn:potential}) is strictly
convex for all $r$ beyond a small cutoff distance for a
large range of parameters. In this study,
we restrict ourselves to such potentials that are convex for $r > 0.1 \sigma$.
\cite{footnote:convexity}

We use an iterative two-step inverse procedure to determine
the optimized parameters of the potential function  (\ref{eqn:potential}) 
under certain constraints that yields the diamond ground state
for a range of pressures.
We define the dimensionless pressure $p^*$ and density $\rho^*$ as follows:
\begin{equation}
p^* = p \sigma^3 / \varepsilon,\qquad \rho^* = \sigma^3 \rho.
\end{equation}
The first step of the optimization procedure involves choosing
an initial set of ``competitor'' configurations. Then we determine
the parameters $a_1$ and $a_2$ that maximize the ratio between the
maximum and minimum pressures $p_{max}$ and $p_{min}$ for which the diamond 
crystal has a lower enthalpy than any competitor configuration, since
we are working in the isobaric ensemble. The second step
involves a rapid cooling procedure within a simulation box
under periodic conditions in the isobaric ensemble, implying
that the simulation box is deforming and changing volume.
We start by choosing initial lattice vectors that define
the simulation box within which there is a  Poisson distributed set of $N$ particles
interacting with the potential function (\ref{eqn:potential}) and the parameters from the first step.
Cooling is then carried out using a quasi-Newton method, which has similar
basins of attraction as obtained from steepest-descent methods or the Metropolis
algorithms at zero temperature. The basis number $N$ for the
periodic cell is varied from 1 to 16
to allow the sampling of variable-basis crystals over the 
entire pressure range defined by $p_{min}$ and $p_{max}$.
If we find a lower enthalpy configuration than that for the
diamond crystal, we add that configuration to the
competitor list and repeat the two-step procedure. 
If we find no other states with lower enthalpy, we terminate the
procedure. Using the final set of competitors, we have access to a large
family of potentials that stabilize the diamond crystal with 
nearly-optimal pressure ranges. This allows us to choose a potential that has
other useful qualities at the cost of only a small 
decrease in the pressure range over which the potential's ground state is 
the diamond crystal. One such desirable property is for the lowest
modes in the phonon spectrum to have the highest possible
energies (\emph{i.e.}, the system is relatively stiff mechanically).

\begin{figure}[htp]
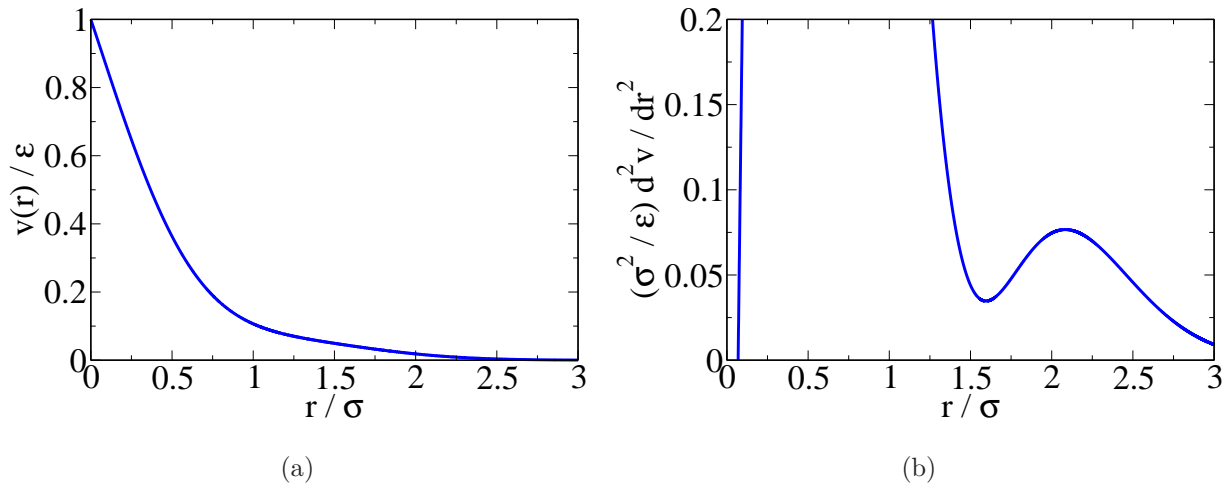

	\centering
	\subfigure[]{
		\includegraphics[scale=0.3]{Fig1a.eps}
		\label{fig:potential}
	}
	\subfigure[]{
		\includegraphics[scale=0.3]{Fig1b.eps}
		\label{fig:sec_deriv}
	}
	\caption{\subref{fig:potential}~Optimized monotonic pair potential $v(r)$ from Eq.~(\ref{eqn:potential}) using the parameters from (\ref{eqn:parameters}): the D1 potential. \subref{fig:sec_deriv}~Second derivative $d^2v/dr^2$ of the same optimized pair potential versus the distance $r$.}
\end{figure}

After optimization under the aforementioned
constraints (\emph{e.g.}, lack of floppy modes and convexity), we obtain
the following optimized parameters for the potential
function (\ref{eqn:potential}):
\begin{equation}
a_1 = -1.42324, \qquad a_2 = 0.713012. \label{eqn:parameters}
\end{equation}
Henceforth, we will refer to the potential function (\ref{eqn:potential}) with
parameters defined by (\ref{eqn:parameters}) as the diamond-1 or D1 potential; 
\cite{footnote:other_params} see Fig.~\ref{fig:potential}.
The diamond crystal is the ground state of the D1 potential from
$p^* = 0.0554$ to $p^* = 0.1010$. At these pressures, the corresponding
densities and nearest-neighbor distances are, respectively, $\rho^* = 0.235$ and $\rho^* = 0.303$, and
$r_{NN} = 1.403 \sigma$ and $r_{NN} = 1.29 \sigma$.
The magnitudes of the second derivative at different distances
have two simultaneous objects: (a) to stabilize the low-coordinated
target structure, and (b) to discriminate against all competitors.
Specifically, using the generalized coordination function formalism described in
Ref.~\onlinecite{Ma11b}, we determined that to obtain low-coordinated ground
states with monotonic convex potentials, the magnitude of the second
derivative must be large
both below and near the nearest-neighbor distance of the low-coordinated
ground state. It also must be small up to
to the nearest-neighbor distance of the close-packed crystals
and large up to the next-nearest-neighbor distance of the targeted
low-coordinated ground state, after which it goes to zero.

We also have attempted to use the same procedure to find a monotonic 
potential of the form (\ref{eqn:potential}) that has the
closely related tetrahedrally-coordinated wurtzite crystal
as its ground state. However, we have found that such a potential does not
exist. This can be readily be explained by the fact that wurtzite and diamond
crystal have the same short-range interparticle distances
(first and second coordination shells), so the only way
a potential of the form (\ref{eqn:potential})
can distinguish between the two is from its longer-range behavior.
However,
potentials of the form (\ref{eqn:potential}) decreases very rapidly at such
distances, since they are dominated by the Gaussian factor.
Therefore, the wurtzite crystal for the potential (\ref{eqn:potential})
has higher energy than the diamond crystal
due to its third nearest neighbors being slightly closer.

\begin{figure}[htp]
	\centering
	\includegraphics[scale=0.4]{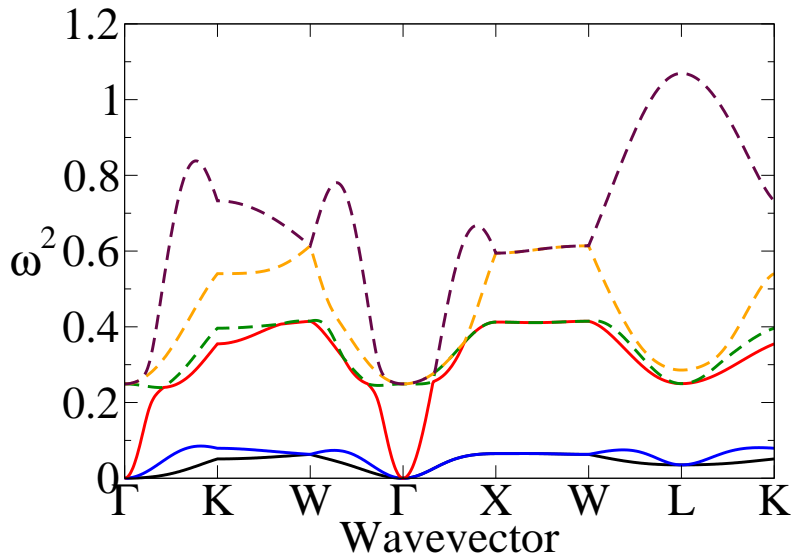}
	\caption{Phonon spectrum in reduced units of the D1 potential for the diamond crystal at dimensionless pressure $p^* = 0.078$ and density $\rho^* = 0.271$. Only a representative subset of wavevectors that lie on paths connecting high-symmetry points ($\Gamma$, $K$, $W$, $X$, and $L$) of the Brillouin zone\cite{As76} is shown. The D1 potential is chosen such that the ratio between the highest and lowest phonon frequencies at the $X$ point is maximized.}
	\label{fig:phonon}
\end{figure}

Figure \ref{fig:phonon} shows
the phonon spectrum (which shows the strengths of the restoring forces for
deformations associated with given wavevectors)
of the D1 potential for the diamond crystal at a pressure
in the middle of its stability range. Full phonon calculations (not only
for the wavevectors shown in Fig.~\ref{fig:phonon}) indicate that the
diamond crystal is indeed mechanically stable for the D1 potential.
The optimized D1 potential is selected 
among those potentials that produce a nearly optimal pressure
range for the diamond ground state in order to achieve a large
ratio between the highest and lowest
phonon frequencies at the $X$ point. While only the $X$ point was considered
in the optimization, the D1 potential is also optimal for the other
wavevectors that we have examined.

\begin{figure}[htp]
	\centering
	\includegraphics[scale=0.7]{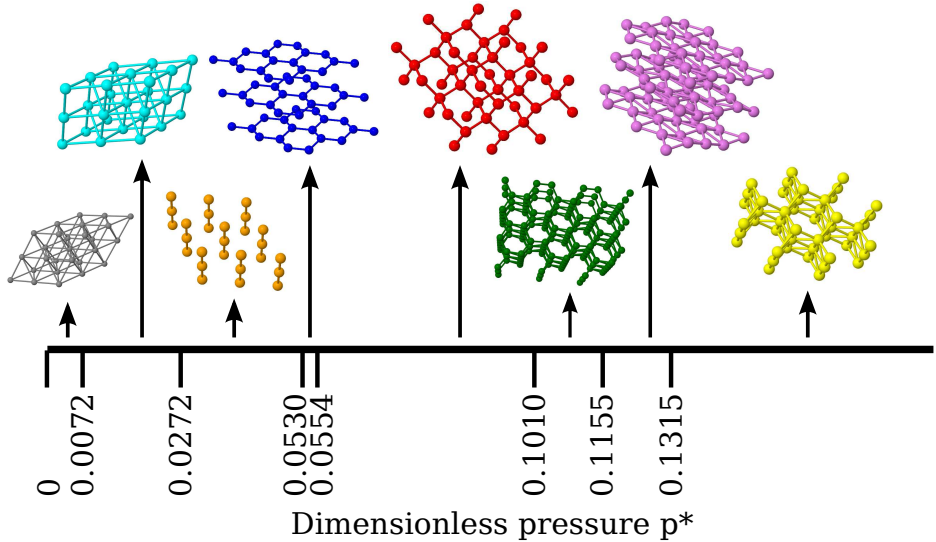}
	\caption{Ground states of the D1 potential for a range of dimensionless pressures obtained from steepest descent for a basis up to $N=16$. The crystal phases indicated from zero pressure to higher pressures are the 12-coordinated face-centered cubic (gray), the 8-coordinated body-centered cubic (cyan), a 2-coordinated hexagonal (orange), a 3-coordinated buckled rhombohedral graphite (blue), the 4-coordinated diamond (red), a 5/6-coordinated deformed diamond (green), a 6-coordinated buckled hexagonal (violet), and a 8-coordinated flattened-hexagonal closed-packed (yellow). Bonds are drawn between particles that are either nearest neighbors or nearly so as to guide the eye. 
}
	\label{fig:phase_diagram}
\end{figure}
The stable phases for the D1 potential at various pressures outside those
for the diamond stability range are shown in Fig.~\ref{fig:phase_diagram}.
The phases are determined  by
repeatedly cooling disordered configurations at constant pressure using the aforementioned
the variable-box energy minimization techniques and 
retaining the resultant configurations whose enthalpy are the lowest.
We find that the diamond crystal is  stable for $0.0554 \le p^* \le 0.1010$. 
Four neighboring phases are particularly interesting. At low pressure,
$0.0272 < p^* < 0.0530$, a hexagonal crystal phase, where the distance
between hexagonal planes is shorter than the distance between particles from
the same plane, is stable. This crystal has an effective coordination number
of two, since the potential favors low-coordinated
configurations over close-packed ones as long as the nearest-neighbor distance
is not too small. Between this phase and the diamond  phase
($0.0530 < p^* < 0.0554$), a low-coordinated rhombohedral graphite
crystalline phase is stable.
The rhombohedral graphite crystal is composed of stacked honeycomb crystals
where each successive honeycomb layer is shifted in the same
direction relative to the layer immediately below it
(unlike standard graphite  where 
the shift direction alternates between layers).
The distance between the planes is $1.5 \sim 1.6$ times larger than the
nearest-neighbor distance within a  layer,
which is much less than that for actual graphite.
At high pressure in the range $0.1155 < p^* < 0.1315$, 
the opposite happens, since the stable phase is a \emph{buckled} simple
hexagonal crystalline, where the nearest-neighbor distance within a hexagonal
plane is shorter than that between planes, resulting in a coordination number
of six. Unlike  the low-pressure hexagonal
crystalline phase, this phase shows buckling: particles in the same layers
are not perfectly aligned, but the distance between nearest neighbors stays
constant.
The transition between the high-pressure buckled hexagonal
phase and the diamond phase ($0.1010 < p^* < 0.1155$) consists
of a highly deformed diamond crystal, for which particles have variable
coordination numbers of either 5 or 6.
The highest-pressure phase shown in Fig.~\ref{fig:phase_diagram} is a
flattened-hexagonal closed-packed crystal, consisting of
a distorted hexagonal closed-packed crystal with layer nearest-neighbor distances
that are larger than the interlayer nearest-neighbor distance. This is not the stable phase for
all $p^* > 0.1315$; other phases arise at higher pressures.

\begin{table}[htp]
	\centering
	\begin{tabular}{|c|c|c|}
		\hline
		$N$ & D1 potential & Star-polymer potential \\
		\hline
		 2 & 96.89\% & 91.41\% \\
		 4 & 89.71\% & 77.38\% \\
		 8 & 62.13\% & 54.28\% \\
		16 & 46.32\% & 26.55\% \\
		32 & 24.57\% &  8.93\% \\
		64 &  5.27\% &  0.30\% \\
		\hline
	\end{tabular}
	\caption{Frequency with which the ground-state diamond crystal is obtained from a steepest descent starting from a random configurations of $N$ particles. For each $N$, the frequency is calculated using 10000 trials, which results in standard deviations smaller than 0.5\%. The D1 potential trials are carried out at $p^*=0.078$, while the star-polymer potential trials used $p / (\frac{5}{18} k_B T f^{3/2})=3.332$ (for which the ground state has a ``packing fraction'' $\eta = 1.2$) and an arm number $f=64$.\cite{Wa99}
}
	\label{tab:succ_rate}
\end{table}

\begin{figure}[htp]
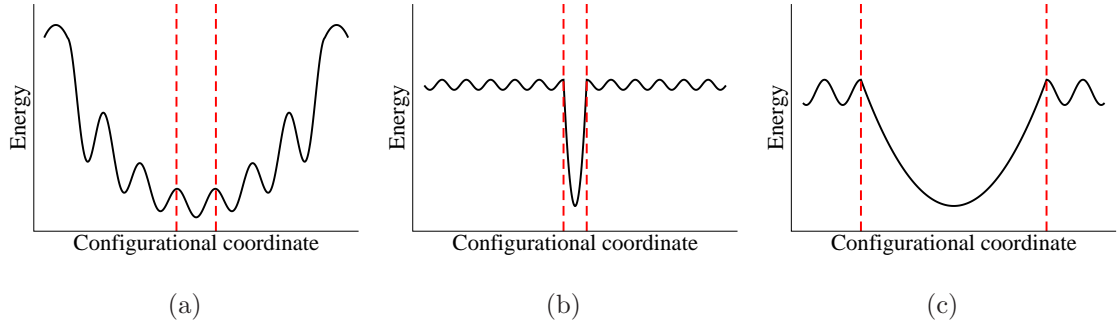

	\centering
	\subfigure[]{
		\includegraphics[scale=0.2]{Fig4a.eps}
		\label{fig:landscape1}
	}
	\subfigure[]{
		\includegraphics[scale=0.2]{Fig4b.eps}
		\label{fig:landscape2}
	}
	\subfigure[]{
		\includegraphics[scale=0.2]{Fig4c.eps}
		\label{fig:landscape3}
	}
	\caption{Schematic representations of three different types of energy landscapes as a function of the configurational coordinate. The boundaries of the basin of attraction associated with the global minima are indicated by the dashed vertical lines. \subref{fig:landscape1}~A relatively rough energy landscape. \subref{fig:landscape2}~An energy landscape with a deep and narrow global minimum. \subref{fig:landscape3}~An energy landscape with a broad and smooth global minimum.}
	\label{fig:landscapes}
\end{figure}

We employed the same rapid cooling method used to map out the phase diagram 
to quantify how easy it is for the system to reach the ground state.
Table~\ref{tab:succ_rate} compares the frequency with which the diamond crystal
ground state is obtained using the D1 and the star-polymer\cite{Wa99} potentials
for various numbers of particles, demonstrating an advantage of the
D1 potential. The high frequency with which the D1 potential results in the
diamond crystal is evidence that its energy landscape is smooth and possesses
a broad global minimum, similar to the schematic illustration in
Fig.~\ref{fig:landscape3}. The
decreasing frequencies for larger $N$ are the consequence of the relative
crudeness of our instant cooling method, which is ineffective at resolving
large-scale defects. However, it is all the more remarkable that this method
is capable of reaching the ground state with reasonable frequency despite
using large bases, as opposed to, for example, a carefully-tuned simulated
annealing procedure.
Nevertheless, we have verified using simulated annealing techniques
with a 256-particle system that the diamond crystal emerges as
the ground state for the D1 potential.

The work reported in this paper provides yet another example of the ``inverse''
statistical mechanical method to identify an appropriate interaction potential
whose non-degenerate classical ground state is a pre-selected crystal
structure.  In general it is not guaranteed at the outset that such a targeted
requirement has a solution.  But in the present case of the fourfold
coordinated diamond lattice previous studies have indeed established that this
can be accomplished with pairwise additive isotropic
potentials.\cite{Re07,Wa99,Pr09}  The existence of these examples establish
that an infinite family of such interactions will produce the diamond
structure as its ground state, each member within some pressure
(\emph{i.e.}, density) range.  

However merely stabilizing a given target structure is typically only part
of the technical objective.  In addition there tend to be properties that one
wishes simultaneously to satisfy or to optimize.  For the present study these
have included maximizing the pressure range of ground-state stability for the
diamond lattice, constraining the potential to monotonicity and convexity,
maximizing the ratio of transverse to longitudinal acoustic sound speeds, and
optimizing capture probability in the desired crystal basin from random
initial configurations.  It should be emphasized that the choice of such
constraints and/or optimizations is not unique, but is driven by overall
scientific objectives.  Distinct choices obviously will identify distinct
optimizing potential functions.
     
The success at constructing diamond potentials naturally raises the question
of whether the structure of that other macroscopic crystalline form of
elemental carbon, graphite, might analogously be the classical ground state
of an isotropic pair potential.  At first sight this might seem easy, given
the existence of potentials that generate the two dimensional analog, the
honeycomb crystal.\cite{Re05} However, the layered structure of this
three-dimensional graphite allotrope, the stable form of elemental carbon
at ambient conditions, with rather large interlayer separation and interlayer
relative shift, realistically appears to present a formidable challenge.
It may turn out that the graphite structure is unattainable with monotonically
decreasing pair interactions.  An ideal circumstance would be to find a
potential whose classical ground state includes both the graphite structure
(at low pressure) and the diamond structure (at elevated pressure), thus
emulating reality.  This ambitious joint requirement might require at
least a combination of two-body and three-body interactions, suggesting a
direction for future research.

At present there is no known constraint on the complexity
(basis of the unit cell) of a single-species
target crystal structure that might be stabilized by an isotropic pair
potential. But as the unit cell of a target crystal structure increases in
size and geometric detail, it is reasonable to suppose that stabilizing
isotropic pair potentials, if they exist, will necessarily also
have to increase in range and complexity. Establishing such a connection
constitutes another direction in which future studies should be focused.

This work was supported by the Office of Basic Energy Science, Division
 of Materials Science and Engineering under Award No. DE-FG02-04-ER46108.
S.T. gratefully acknowledges the support of a Simons Fellowship in Theoretical
Physics, which has made his sabbatical leave this entire academic year
possible.


\begin{thebibliography}{26}
\bibitem{forward}
The forward approach specifies interparticle interactions 
for a many-particle system and then studies 
structural, thermodynamic and kinetic features of the system.
\bibitem{private}
Private communication. Phonon spectra were not provided in Ref. \onlinecite{Wa99}.
\bibitem{footnote:convexity}
Since the nonconvexity of the potential is restricted to very
small distances, one
can add a short-range hard-core repulsion to Eq.~(\ref{eqn:potential}) to
obtain a convex potential, which will behave exactly like the
original potential, provided that the pressure of temperature
are not extremely large.
\bibitem{footnote:other_params}
For example, we denote by D2 the potential that has the 
largest ratio of maximum to minimum  stable  pressures
for the diamond ground state.
Its parameters are $a_1 = -1.21189$, and $a_2 = 0.788955$.
This potential possesses softer transverse phonon modes compared
to the D1 potential.
\end{thebibliography}
\end{document}